\documentclass[conference]{IEEEtran}
\IEEEoverridecommandlockouts
\usepackage{cite}
\usepackage{amsmath,amsfonts,amssymb,amsthm}
\usepackage{algorithmic}
\usepackage{graphicx}
\usepackage{textcomp}
\usepackage{xcolor}
\usepackage[utf8]{inputenc}
\usepackage{commath}
\usepackage{scalerel}
\usepackage{tikz}
\usepackage[]{changes}

\usepackage{pgfplots}
\usepackage[hidelinks]{hyperref}
\pgfplotsset{
    compat=newest,
    every axis/.append style={font=\scriptsize,
                              width=\linewidth,
                              height=0.6\linewidth}}

\pgfplotsset{compat=1.14,
    /pgfplots/ybar legend/.style={
    /pgfplots/legend image code/.code={%
       \draw[##1,/tikz/.cd,yshift=-0.25em]
        (0cm,0cm) rectangle (3pt,0.8em);},
   },
}
\usepgfplotslibrary{groupplots}
\usepgfplotslibrary{dateplot}

\usetikzlibrary{svg.path}

\newcommand\orcidicon[1]{\href{https://orcid.org/#1}{
\includegraphics[height=0.8\baselineskip]{orcid.ps}
}}

\renewcommand{\del}[1]{}

\def\BibTeX{{\rm B\kern-.05em{\sc i\kern-.025em b}\kern-.08em
    T\kern-.1667em\lower.7ex\hbox{E}\kern-.125emX}}

\begin{document}

\title{Fixed mmWave Multi-User MIMO: Performance Analysis and Proof-of-Concept Architecture\\
\thanks{This research is supported by the Research Foundation Flanders (FWO), project no. S003817N (OmniDrone) and by the European Union’s Horizon
2020 under grant agreement no. 732174 (ORCA project) .}
}

\author{\IEEEauthorblockN{
Achiel Colpaert, 
Evgenii Vinogradov, 
and Sofie Pollin
}
\IEEEauthorblockA{KU Leuven, ESAT - Department of Electrical Engineering, Kasteelpark Arenberg 10, 3001 Heverlee, Belgium\\
E-mail: \{achiel.colpaert, evgenii.vinogradov, sofie.pollin\}@kuleuven.be}}

\maketitle

\begin{abstract}
In this paper, we present a fixed mmWave Multi-User Multiple-Input Multiple-Output (MIMO) system for fixed wireless access with a unique architecture. A digital MIMO system is combined with an analog multi-beam antenna array which uses a high-dimension 16x16 Butler matrix to obtain 16 orthogonal beams. A system model of this architecture is presented and used to simulate its performance comparing to the performance of common-used patch antennas. Several MIMO precoding techniques are considered and compared with basic analog beamforming. To verify these results, a prototype is built and a dedicated measurement campaign is performed. The results show that the system model is a good approximation and that the use of the multi-beam antenna array is a good alternative to patch antennas for a large number of users.
\end{abstract}

\begin{IEEEkeywords}
5G, mmWave, multi-user, MIMO, beamforming, butler matrix, spectral efficiency
\end{IEEEkeywords}

\section{Introduction}
The release of 5G New Radio defines a three times higher spectral efficiency and a hundred times better energy efficiency compared to 4G. It also specifies large downlink data-rates for users in dense urban areas while also providing good coverage\cite{outlook}. One of the enablers for achieving this ambitious performance is the usage of spectrum in higher frequencies, so-called millimeter wave (mmWave) frequencies, combined with various beamforming techniques.

For example, broadband access at home is traditionally  provided by wired connections i.e., fiber or copper. However, in rural areas, wired access becomes difficult due to the large area to cover resulting in a high deployment cost. This problem contributes to the increasing digital divide between rural and urban areas, especially in developing countries. In this context, 5G shows a promising alternative by providing wireless broadband access at a reduced cost, bringing high data rates to the home \cite{outlook}.

In this work, Fixed Wireless Access is considered as the first main application for high throughput mmWave links. In this case, the mmWave link is used to bridge the last meters between the optical fiber termination point and a fixed Customer Premise Equipment in a home. A single mmWave transmission head in the street can serve multiple homes, thus it becomes interesting to study the feasibility of low-cost spatial multiplexing. Every home will be able to benefit from a dedicated mmWave link, achieving several Gbps over its downlink in a dedicated and exlusive fashion.

Exploiting the 26~GHz to 300~GHz mmWave frequency bands allows for the use of high bandwidths and data rates. However, mmWave also brings difficulties due to higher propagation losses and blockages \cite{mmwave-problems,Achiel18}. To compensate these higher losses, both analog and digital beamforming, techniques have been suggested \cite{mmwave-problems,Achiel18}. Analog beamforming allows for low cost, low profile hardware but offers limited flexibility. On the contrary, digital beamforming allows for high flexibility and spatial multiplexing but it is expensive and has a high energy cost. To combine the best of both worlds, several hybrid beamforming systems have been proposed in literature in an effort to reduce to energy cost but simultaneously keep the flexibility. Authors of \cite{hybrid-beamforming-survey} give an extensive overview of all the published work in this context. In \cite{hybrid-feasibility}, a feasibility study and prototype of a hybrid beamforming system is given. A realtime Software Defined Radio (SDR) testbed is used to build this prototype.

In this work, we focus on a digital Multi-User Multiple-Input Multiple-Output (MIMO) base station (BS) in combination with a fixed analog phase shifting antenna array.
Contributions of this paper:
\begin{itemize}
\item A Fixed Multi-User MIMO architecture and proof-of-concept;
\item Comparison of beam selection and beam precoding using both simulation and experiment;
\item Evaluation of fixed precoding techniques using over-the-air experiments and simulations, which are based on both radiation patterns and user locations.
\end{itemize}
The rest of this paper is organized as follows. The equipment used in the system is described in Section \ref{sec:equipment}. The system model is introduced in Section \ref{sec:system_model}. Followed by Section \ref{sec:results}, which shows the simulation and measurement results. Finally, the conclusion is drawn in Section \ref{sec:conclusion}.

\section{Equipment and Scenario}\label{sec:equipment}
In this section, we discuss the equipment used in the experimental setup. First, we shed some light on the multi-beam antenna array followed by a short description of the MIMO testbed.
\subsection{Antenna Array}\label{sec:array}
\begin{figure}[tbp]
\centerline{\includegraphics[width=0.9\linewidth]{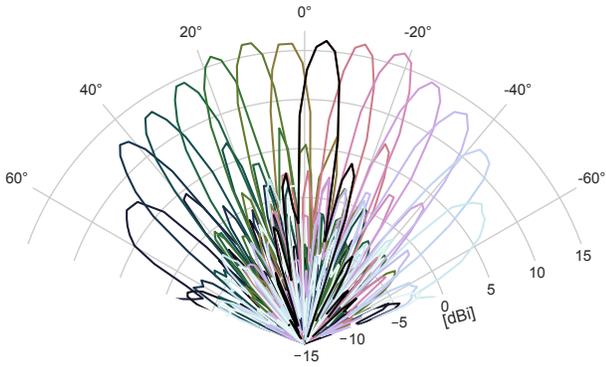}}
\caption{Measured azimuth beam pattern of the beam-forming antenna array. Every color represents a different beam.}
\label{fig:pattern}
\end{figure}

The multi-beam antenna array used in this paper is an extended prototype of the work in \cite{antenna}. It features a 16x16 Butler-matrix (BM) as phase shifting network connected to a 1x16 linear antenna array. The linear antenna array is implemented on the same printed circuit board as the BM and consists of quasi-Yagi antenna elements. The operating frequency of this multi-beam array is $25$~to~$30$~GHz, while the input intermediate frequency (IF) is $2.4$~GHz, up- and down-conversion is handled by the multi-beam array. This IF allows the use of cheaper off-the-shelf devices omitting expensive mmWave equipment. The BM is capable of generating spatial orthogonal beams, resulting in every beam having nulls in the directions of the main lobes of the other beams. The half power beam width of these beams is $7$~deg, the maximum beam gain is $16$~dBi and the total spatial angle range is $\pm68$~deg. The measured antenna pattern can be seen in Fig.~\ref{fig:pattern}.
 Each of the 16 beam directions of the multi-beam array has a corresponding RF input/output port at the IF resulting in a fully connected 16x16 structure between the inputs ports and the antennas.

\subsection{Testbed}
The KU Leuven Massive MIMO testbed, based on LTE-TDD, is used, described in detail in \cite{mimo-basestation}. The testbed is configured to use $2.4$~GHz as its center frequency, which is the IF of the antenna array, and uses a $20$~MHz bandwidth. The system uses OFDM modulation with a total of 1200 subcarriers. There are two main components to this testbed, the first being the BS, which in this setup has 16 RF-chains connected to the 16 inputs of the multi-beam array. The second component are the users, in this paper we use one user and virtually add more users offline. Both components run the LabVIEW Communications MIMO Application Framework \cite{mimo-framework}. The RF chain of the user is connected to a 16-port switch \cite{rf-switch}, which can be manually controlled over a control interface by the user. Both antenna arrays are synchronised using a $11.8$~GHz clock generated by an ERASynth+ RF Signal Generator \cite{erasynth}.

\section{System Model}
\label{sec:system_model}
\begin{figure}[tbp]
\centerline{\includegraphics[width=1\linewidth]{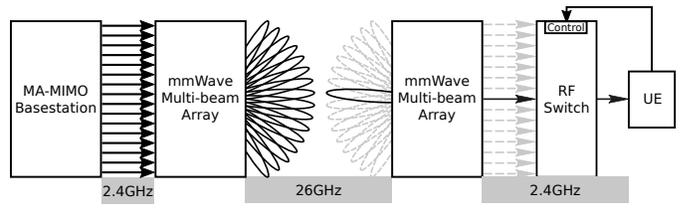}}
\caption{Schematic illustration of the system indicating the interconnections between hardware components as well as the used frequencies.}
\label{fig:schematic}
\end{figure}

In this work, we consider a fixed downlink transmission in a single-cell multi-user mmWave MIMO system. The system is equipped at both sides with the multi-beam array described in Section \ref{sec:array} and is capable of transmitting $K \leq 16$ spatial data streams serving $K$ users simultaneously.  A schematic overview of the BS and user can be seen in Fig.~\ref{fig:schematic}.

\par
The BS uses a baseband digital precoder defined as follows
\begin{equation}
    \mathbf{F}^{BB} = [\mathbf{f}^{BB}_{1},\dots,\mathbf{f}^{BB}_{K}],
    \label{eq:FBB}
\end{equation}
where $\mathbf{F}^{BB} \in \mathbb{C}^{16\times K}$, $\mathbf{f}^{BB}_{k}$ is the precoder for the ${k}$-th user and $\norm{\mathbf{f}^{BB}_{k}}=1$. These precoding vectors are calculated based on the channel estimation in the uplink. The precoding vector calculation is described later in Section~\ref{sec:precoder}.

The transmitted signal vector is given by
\begin{equation}
    \mathbf{s} = \mathbf{F}^{BB}\mathbf{x},
    \label{eq:s}
\end{equation}
where $\mathbf{s}\in \mathbb{C}^{16\times 1}$, and $\mathbf{x} = [x_{1},\dots,x_{K}]^T$ contains the symbols for $K$ users.

\subsection{Signal Model}
Consider a discrete memoryless channel where the received signal vector for a single user $k$ is given by
\begin{equation}
    \mathbf{y}_{k} = \sqrt{p_{tx}}\mathbf{H}_{k}\mathbf{s} + \mathbf{n}_{k},
\label{eq:yk}
\end{equation}
where $\mathbf{y}_{k} = [y_{1,k},\dots,y_{16,k}]$ contains the received signals at the 16 receive antennas, $p_{tx}$ is the transmit power at the BS, $\mathbf{s}$ is the transmitted signal vector as defined in \eqref{eq:s} and $\mathbf{H}_{k} \in \mathbb{C}^{16\times 16}$ is the MIMO channel impulse response (CIR) matrix between the BS and user $k$ (see Section~\ref{sec:H}). Readers should take note that this CIR matrix does include the multi-beam array impulse responses at both BS and UE side. $\mathbf{n}_{k} \sim \mathcal{NC}(0,\,\sigma^{2}) \in \mathbb{C}^{16 \times 1}$ is the normalized Gaussian noise vector at the receiver of user $k$, where $\sigma^2$ is the noise power.
Substituting \eqref{eq:s} in \eqref{eq:yk} gives the following received signal vector:
\begin{equation}
    \mathbf{y}_{k} = \sqrt{p_{tx}}\mathbf{H}_{k}\mathbf{F}^{BB}\mathbf{x} + \mathbf{n}_{k}.
    \label{eq:yk2}
\end{equation}
\par
To receive the signal, the user selects the beam detecting the highest power using the RF switch. For user $k$, this switch is represented by the vector $\mathbf{w}^{RF}_k \in \mathbb{Z}_2^{16 \times 1}$, which contains only one non-zero element dependent on the optimal beam.
 The received signal $y_{n,k}$ at the user $k$ after $n$-th antenna selection can then be represented as follows:
\begin{equation}
    y_{n,k} = (\mathbf{w}^{RF}_k)^T\mathbf{y}_{k}.
\end{equation}
Adding the analog combining vector to the received signal vector in \eqref{eq:yk2} gives us the final expression for the received signal at the user $k$:
\begin{equation}
    y_{n,k} = \sqrt{p_{tx}}(\mathbf{w}_k^{RF})^{T}\mathbf{H}_{k}\mathbf{F}^{BB}\mathbf{x} + (\mathbf{w}_k^{RF})^T\mathbf{n}_{k}.
\end{equation}

\subsection{Fixed LoS channel model}\label{sec:H}
The user and BS are in line of sight (LoS).  The channel is modeled for one subcarrier resulting in a narrowband frequency flat channel. All users are located at fixed locations with a relative angle to the BS of $\theta_{bs}$. The users are rotated with an angle $\theta_k$ towards the BS, resulting in the following deterministic channel response for user $k$ \cite{massivemimobook}:
\begin{equation}
    \mathbf{H}_k(\theta_{bs},\theta_k) = \sqrt{\mathbf{B}(\theta_{bs}, \theta_k)}\circ(\Phi_{rx}(\theta_k)\Phi_{tx}^T(\theta_{bs})),
    \label{eq:Hk}
\end{equation}
where $\circ$ represents the Hadamard product, $\Phi_{rx}(\theta_k) \in \mathbb{C}^{16 \times 1}$ and $\Phi_{tx}(\theta_{bs}) \in \mathbb{C}^{16 \times 1}$ represent the $\theta_k$- and $\theta_{bs}$-dependent phase shifts seen at the receiver and transmitter ULAs, respectively. These phase shifts represent both the fixed phase shifts in the BM and the angle dependent phase shifts between the antenna array elements. $\mathbf{B}(\theta_{bs},\theta_k) \in \mathbb{C}^{16 \times 16}$ represents the angle dependent channel gain matrix and can be interpreted as the macroscopic large-scale fading between all transmitting and receiving antennas. $\mathbf{B}(\theta_{bs},\theta_k)$ also contains the antenna gains, which results in its dependency on $\theta_{bs}$ and $\theta_k$. Thus, $\mathbf{B}_k(\theta_{bs}, \theta_k)$ can be written as
\begin{equation}
    \mathbf{B}_k(\theta_{bs}, \theta_k) =\mathbf{g}^{rx}(\theta_k)\cdot(\mathbf{g}^{tx}(\theta_{bs}))^{T}\cdot PL_k\cdot L_k,
    \label{eq:Bk}
\end{equation}
where $L_k$ are the cable and other hardware related losses, $PL_k$ is the free space path loss dependent on the distance between user and BS. Where $\mathbf{g}^{tx}(\theta_{bs}) = [g^{tx}_{1}(\theta_{bs}), g^{tx}_{2}(\theta_{bs}), \dots, g^{tx}_{16}(\theta_{bs})]^T \in \mathbb{C}^{16 \times 1}$ is the transmitter antenna gain vector towards user $k$. Each element $g^{tx}_{m}(\theta_{bs})$ represents the gain of antenna $m$ towards user $k$ dependent on the azimuth angle $\theta_{bs}$ of the user to the BS. Similarly, $\mathbf{g}^{rx}(\theta_k) \in \mathbb{C}^{16 \times 1}$ consists of the $\theta_k$-dependent antenna gains of the receiver multi-beam array.
\par
To increase further readability, we omit the $\theta_{bs}$ and $\theta_k$ dependency in the notation of the channel impulse response matrix and shorten it to $\mathbf{H}_k$. As the user $k$ uses only one of its receiving antennas, we can simplify the channel matrix $\mathbf{H}_k$ to a channel vector $\mathbf{h}_k$ by applying the user's combining vector $\mathbf{w}^{RF}_k$. This vector $\mathbf{h}_k$ contains only one antenna beam of the user side multi-beam antenna array. The simplified channel vector $\mathbf{h}_k \in \mathbb{C}^{16 \times 1}$ is then defined as follows
\begin{equation}
    \mathbf{h}_k = ((\mathbf{w}^{RF})^T\mathbf{H}_k)^T.
    \label{eq:hk}
\end{equation}
Take note that this is the format of the channel that the BS will retrieve from its channel estimation.

\subsection{Digital Precoding}\label{sec:precoder}
For the calculation of the digital precoder at the BS, the CIR is necessary. The CIR can be gathered at the BS side by several different channel estimation techniques, for example using minimum mean squared error channel estimation defined in Theorem 3.1 in \cite{massivemimobook}. However, these estimated channels do contain estimation errors and are thus indicated by $\mathbf{\hat{h}}_k$ to differentiate them from the real channel $\mathbf{h}_k$.
\par
In this paper, we consider three digital precoding techniques, namely Maximum Ratio (MR) precoding, Zero Forcing (ZF) and Regularized Zero Forcing (RZF). These techniques are based on their corresponding combining matrices as defined in \cite{massivemimobook}. The precoding vector for the $k$-th user can be calculated as follows:
\begin{equation}
    \mathbf{f}_k = \frac{\mathbf{w}^{BB}_k}{\norm{\mathbf{w}^{BB}_k}},
\end{equation}
where $\mathbf{w}^{BB}_k$ is one vector of the digital combining matrix at the BS: $\mathbf{W}^{BB}=[\mathbf{w}^{BB}_1,\dots,\mathbf{w}^{BB}_k]$. This combining matrix is calculated using the CIR of all users: $\mathbf{\hat{H}} = [\mathbf{\hat{h}}_1, \dots,\mathbf{\hat{h}}_k] \in \mathbb{C}^{16 \times K}$. A short overview of these three techniques is given, for a more detailed description we refer to \cite{massivemimobook}:

\begin{itemize}
    \item \textit{MR} maximizes the received signal to all users, $\mathbf{W}^{BB} \in \mathbb{C}^{16 \times K}$ is defined as follows
\begin{equation}
    \mathbf{W}^{BB,MR} = \mathbf{\hat{H}}.
\end{equation}
This precoding technique is often used due to its low complexity.
    \item \textit{ZF}  attempts to cancel all intra-cell interference. The combining matrix is defined as
\begin{equation}
    \mathbf{W}^{BB,ZF} = \mathbf{\hat{H}}((\mathbf{\hat{H}})^H\mathbf{\hat{H}})^{-1}.
\end{equation}
\item \textit{RZF}  makes a trade-off between noise and intra-cell interference. The combining matrix is defined as follows
\begin{equation}
    \mathbf{W}^{BB,RZF} = \mathbf{\hat{H}}((\mathbf{\hat{H}})^H\mathbf{\hat{H}}+\sigma^2\mathbf{P}_{tx}^{-1})^{-1},
\end{equation}
where $\mathbf{P}_{tx} = diag(p_{tx},\dots,p_{tx}) \in \mathbb{C}^{16 \times 16}$ is a diagonal matrix containing the transmit power to each user and $\sigma^2$ is the noise power.
\end{itemize}
\subsection{Spectral Efficiency}
To verify the performance of the proposed system, we evaluate the downlink spectral efficiency (SE) expressed in bps/Hz.
A single cell system is considered and the individual downlink SE for user $k$ is as follows
\begin{equation}
    SE^{DL}_k = \frac{\tau_d}{\tau_c}\log_2(1 + SINR^{DL}_k),
\end{equation}
where $SINR^{DL}_k$ is the downlink Signal-to-Interference-and-Noise-Ratio (SINR) for the user $k$ and $\frac{\tau_d}{\tau_c}$ is the fraction of samples per coherence block that is used for downlink data. The $SINR^{DL}_k$ is defined as follows:
\begin{equation}
    SINR^{DL}_k = \frac{p_{tx}|(\mathbf{w}^{RF}_k)^T\mathbf{H}_k\mathbf{f}^{BB}_{k}|^2}{\sum_{i\neq k}^{K}p_{tx}|(\mathbf{w}^{RF}_k)^T\mathbf{H}_k\mathbf{f}^{BB}_{i}|^2+\sigma^2_{DL}},
\end{equation}
where  $\mathbf{H}_k$ is the real channel matrix for user $k$ as defined in \eqref{eq:Hk}, $\mathbf{f}^{BB}_k$ the digital baseband precoding vector for user $k$ as defined in \eqref{eq:FBB} and $\sigma_{DL}^2$ being the average noise power.
\par
We can also evaluate the summed SE per cell, expressed in bps/Hz/cell. Which is calculated by summing the individual SE of all users within one cell:
\begin{equation}
    SE^{DL}_{sum} = \sum_{k = 1}^{K}SE^{DL}_k.
    \label{eq:sum}
\end{equation}

\section{Results}\label{sec:results}
To evaluate the performance of this system, a set of simulations was performed, followed by a dedicated measurement campaign to verify these simulation results.
First, we discuss the used parameters, followed by an overview of the simulation and measurement methods. Finally, a short interpretation of the results is given.
\par
The parameters of the simulation are determined by the scenario of the measurements. We consider the number of users ranging from $1$ to $16$. We use an operating frequency of $26$~GHz and a signal bandwidth of $20$~MHz. This bandwidth is determined by our experimental setup using an LTE based framework. For each user $k$, we generate a channel $h_k$ using equation \eqref{eq:hk}. To each of these channels we add noise $N$ to simulate the estimated channel $\hat{h}_k$. Noise power is calculated with the expression $N=N_0BF$, where $N_0$ is the noise density, $B$ is the signal bandwidth and $F$ is the noise figure of the UE. We assume a random azimuth angle $\theta_{bs}$  of $\pm60$~deg according to a uniform distribution, where $0$~deg is perpendicular to the ULA. For each angle the transmit antenna gain vector $\mathbf{g}^{tx}$ is obtained from the measured antenna pattern. The rotation of the user $\theta_k$ is fixed to 4 degrees as we assume the user will be able to align one of its beams, thus, the antenna gain $\mathbf{g}^{rx}$ is fixed to $16$~dBi. The users are located at a fixed distance of $5$ meters from the BS, which is imposed by our experimental setup.  The values of all the other parameters used in the simulation setup can be found in Table~\ref{tab:parameters}.

\begin{table}[tbp]
\caption{Simulation parameters}
\begin{center}
\begin{tabular}{|c|c|}
\hline
\textbf{Parameter} & \textbf{Value}       \\
\hline
\hline
number of users, $K$        & $1$ to $16$   \\
base station angle, $\theta_{bs}$      & $\pm60$ deg\\
user angle, $\theta_{k}$      & $4$ deg\\
center frequency, $f$       & $26$ GHz      \\
signal bandwidth, $B$       & $20$ MHz      \\
transmit power, $p_{tx}$    & $3$ dBm       \\
distance, $d$               & $5$ m         \\
noise density, $N_0$          & $-174$ dBm/Hz        \\
noise figure, $F$               & $9$ dB         \\
$\tau_d/\tau_c$     & $1$ \\
number of realisations, $I$        & $1000$ \\
\hline
\end{tabular}
\label{tab:parameters}
\end{center}
\end{table}

The proposed system is benchmarked against a simulated linear array of half-wavelength rectangular patch antennas. Each simulation consists of generating a set of $K$ different user positions and calculating the $SE_k^{DL}$ by averaging over $100$ channel estimations for each user. For statistical analysis, we perform $I=1000$ realisations of this process for each number of users $K$.
\par
Corresponding measurements are performed using the equipment described in Section \ref{sec:equipment} and using the same parameters as in Table \ref{tab:parameters}. In an indoor environment, we put the BS and UE at a fixed distance $d$ and rotated the UE $\theta_k$ degrees. We performed a channel estimation at 150 random rotation angles $\theta_{bs}$ of the BS. The system performs 100 channel estimations per angle. During offline processing, these channel estimations are regarded as the real channel $h_k$ and noise $N$ is added on top of it to represent the estimated channel $\hat{h}_k$. For each number of users $K$, we also generate $I=1000$ different realisations by selecting $K$ random rotation angles for each realisation. The SE of each UE in a realisation is calculated for all $100$ channel estimations and then averaged to get the user's expected SE. Within one realisation, the summed SE is obtained using equation~(\ref{eq:sum}).
\par
Analog precoders are generated to compare the proposed system to a system without MIMO precoding. This can be achieved by generating a precoding vector where all the elements are zero except the element of the beam directed towards the user. Doing this, we simulate an analog beamforming system such as the one at the user side.
\par
Next, we evaluate both the individual and summed spectral efficiency in function of the number of users $K$.
\subsection{Individual Spectral Efficiency}
\begin{figure}[tbp]
\centerline{
\includegraphics{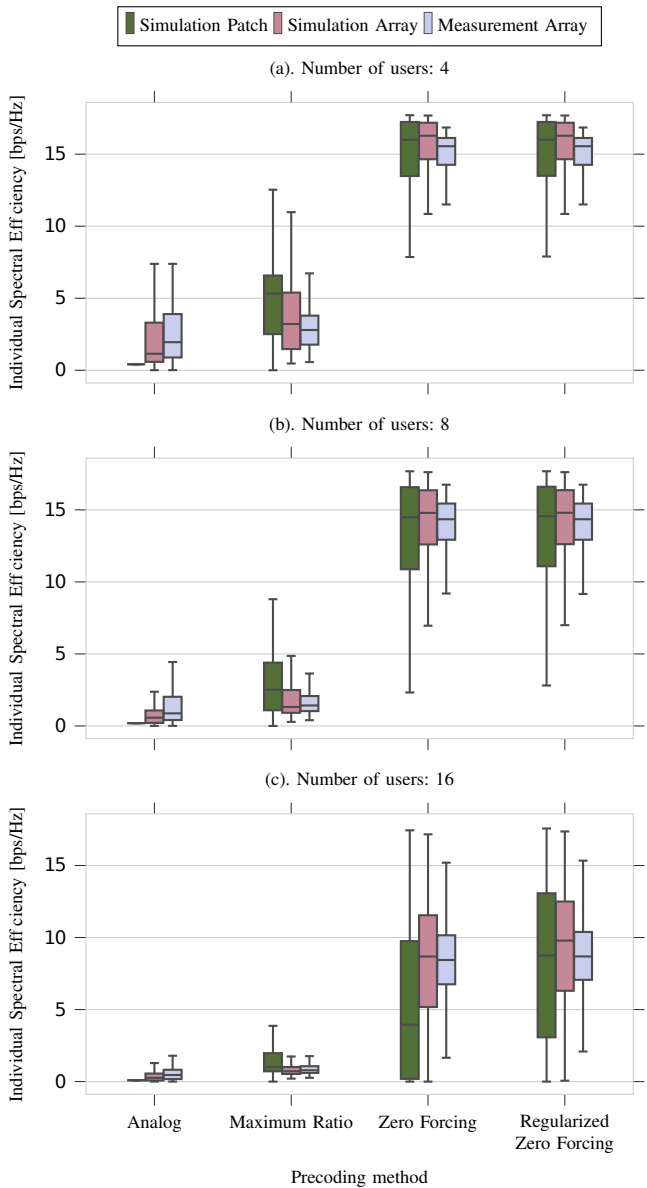}
}
\caption{Individual spectral efficiency for different number of users showing a comparison between different precoding techniques as well as a comparison between the simulated and measured data.}
\label{fig:individ_se}
\end{figure}

The individual spectral efficiency can be seen in Fig. \ref{fig:individ_se}. The median value gives an indication of the average performance of the system while the distance between the lower and the upper quartile gives an indication of the fairness of the system. A fair system has a small spread of its individual SE.
\par
When using MR precoding, the patch antenna array always performs better in terms of individual SE, thus we can conclude that a classic array of patch antennas is the most efficient option when MR precoding is considered. However, both ZF and RZF generally outperform the Analog and MR precoders and they have similar performance for a low amount of users for both arrays. However, for a number of users equal to the number of antennas, ZF's performance starts to degrade, especially if we consider fairness. The measured system is almost always fairer than the simulated system, except for analog beamforming. ZF and RZF perform similar for the beamforming array, this could be due to the limited noise power. If a larger bandwidth is considered, then RZF should outperform ZF due to its noise compensation. We can conclude that the patch and beamforming antenna arrays are both efficient for low number of users. However, if we want to serve a high number of users similar to the number of antennas, then a directive antenna array proves more efficient and fair, ideally using RZF precoding.
\par

\subsection{Summed Spectral Efficiency}
\begin{figure}[tbp]
\centerline{
\includegraphics{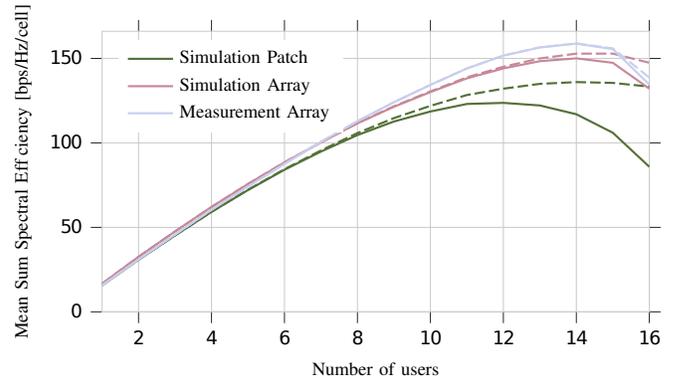}
}
\caption{Summed spectral efficiency compared to the number of users. The solid and dashed lines represent ZF and RZF precoding, respectively.}
\label{fig:sum_se}
\end{figure}

When the summed SE is considered for analog and MR precoding, we can draw similar conclusions as when individual SE is analyzed. The only difference is that the system performance is independent of the number of users.
Fig.~\ref{fig:sum_se} shows the mean summed SE per cell for different numbers of users using ZF and RZF. The patch and beamforming antenna arrays show a similar performance for up to  eight users. For a larger number of users, the patch array's performance degrades faster than the beamforming antenna array's for both ZF and RZF. This confirms the conclusion that for a number of users similar to the number of antennas, the beamforming antenna array outperforms the patch antenna array. The performance of the simulated antenna array approximates on average the measured results. This verifies that the proposed model accurately reproduces the behavior of the real system.

\par

\section{Conclusion}\label{sec:conclusion}
This paper presents a mmWave MIMO system prototype with a unique architecture for fixed wireless access. An analog multi-beam antenna array, realised with a BM, is combined with a digital MIMO system. First, this architecture is benchmarked against a standard patch antenna array. Followed by a comparison of both analog and digital precoding techniques. These results are then verified by a dedicated measurement campaign. It is concluded that the patch antenna array performs similar to the beamforming array up to a limited amount of users. Once the number of users reaches the number of antennas, the beamforming array outperforms the patch array in terms of spectral efficiency and fairness. It is also concluded that the proposed model of the beamforming array is a good approximation of the hardware. However, a more extensive measurement campaign is necessary to further evaluate the model and to test the prototype to its full extent. Both the model and propotype give us insight into developing for future mmWave MIMO communication systems and can be used to further explore other topics in this domain, i.e., user mobility.

\bibliography{biblio}{}
\bibliographystyle{IEEEtran}
\end{document}